\magnification=\magstep1 \overfullrule=0pt
\parskip=6pt
\baselineskip=15pt
\headline={\ifnum\pageno>1 \hss \number\pageno\ \hss \else\hfill \fi}
\pageno=1
\nopagenumbers
\hbadness=1000000
\vbadness=1000000

\input epsf

\vskip 25mm
\vskip 25mm
\vskip 25mm

\centerline{\bf A NOTE ON A THEOREM OF BOURBAKI } \vskip 10mm

\centerline{\bf Hasan R. Karadayi} \centerline{Dept. Physics, Fac.
Science, Istanbul Technical University} \centerline{34469, Maslak,
Istanbul, Turkey } \centerline{e-mail: karadayi@itu.edu.tr}

\vskip 25mm

\centerline{\bf{Abstract}}

We have recently show that Poincare series of Hyperbolic Lie
algebras have the form of a ratio between Poincare series of a
chosen finite Lie algebra and a polynomial of finite degree. By the
aid of some properly chosen examples, we now give some remarks on a
related theorem of Bourbaki.

\hfill\eject

\vskip 3mm \noindent {\bf{I.\ INTRODUCTION }} \vskip 3mm

Let $ P(G_N) = \sum_s d(s) \ t^s  $ be the Poincare Series of a
Kac-Moody Lie algebra $G_N$ of rank N. Here and in the following, t
denotes an indeterminate. It is known that $d(s)$ is the number of
Weyl group elements which are composed out of the products of s
number of simple Weyl reflections corresponding to simple roots of
$G_N$ {\bf [1]}. In this work, this will be adopted as the
definition of Poincare series of Kac-Moody Lie algebras.

For finite Lie algebras, Poincare polynomials are known in the
following form

$$  P(G_N) = \prod^N_{i=1}  { t^{\nu_i}-1 \over t-1}  \eqno(I.1) $$

\noindent where $\nu_i$'s are the degrees of N basic invariants of
$G_N$. For an affine Kac-Moody Lie algebra $\widehat{G}_N$
originated from a generic finite Lie algebra $G_n$ \ ( $N>=n$ in
general), Bott theorem {\bf [2]} states that its Poincare polynomial
has the following product form

$$ P(\widehat{G}_N)=P(G_n) \prod^N_{i=1} {1 \over 1-t^{\nu_i-1}} \ .  \eqno(I.2) $$

To the author's knowledge, explicit results seem to be lacking
beyond affine ones although there is a general theorem which is
known to be valid in any case. It is this theorem{\bf [3]} which in
fact can be applied also in obtaining of (I.1) and also (I.2).
Theorem states that
$$P(G_N)=P(g_n) \ R  \eqno(I.3) $$
\noindent where $P(g_n)$ is Poincare polynomial of a sub-algebra
$g_n\subset G_N $ with the trivial condition that $g_n$ should be
contained inside the Dynkin diagram of $G_N$. A non-trivial
assertion of theorem is that R is a {\bf rational function}.

For a Hyperbolic Lie algebra $ H $, our observation is that its
Poincare polynomial comes in the form
$$  P(H)={P(g) \over Q(g) }   \eqno(I.4) $$
\noindent where $g$ is a properly chosen finite Lie Algebra and
$Q(g)$ is a polynomial of some finite degree in indeterminate t. Due
to the fact that a rational function or its inverse can only be
represented by a polynomial of infinite order, (I.3) and (I.4) seem
to be in contradiction. This will be exemplified in the following.

We know only a finite number of Hyperbolic Lie algebras {\bf [4]}.
Let us proceed in the concrete example for H with the Dynkin
diagram:

\epsfxsize=6cm \centerline{\epsfbox{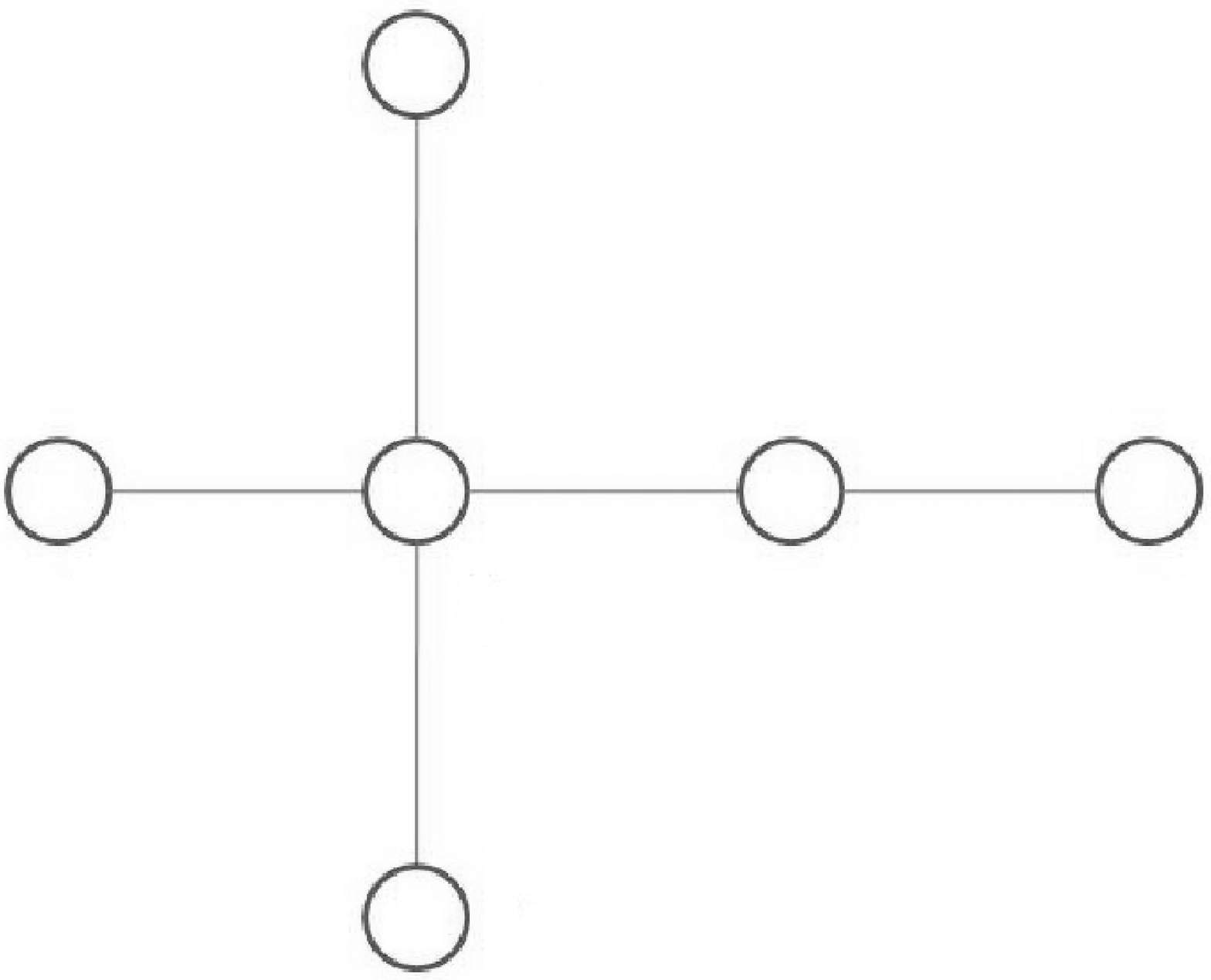}}

\noindent Following 3 examples will be instructive in applications
of above mentioned theorem.

Let $W({H})$ be the Weyl group of H and $ \sigma_i$'s be its
elements corresponding to simple roots $ \alpha_i$'s where $i=1,
\dots , 6$ . We assume that reduced Weyl group elements can be
expressed by
$$ \Sigma(i_1, \dots ,i_k) \equiv \sigma_{i_1}. \dots \sigma_{i_k}. \eqno(I.5) $$
Let $ S=\{\alpha_1, \dots , \alpha_6\} $ be the set of simple roots
of $H$ for which we use following 2 enumerations of its Dynkin
diagram for the first 2 examples:

\epsfxsize=6cm \centerline{\epsfbox{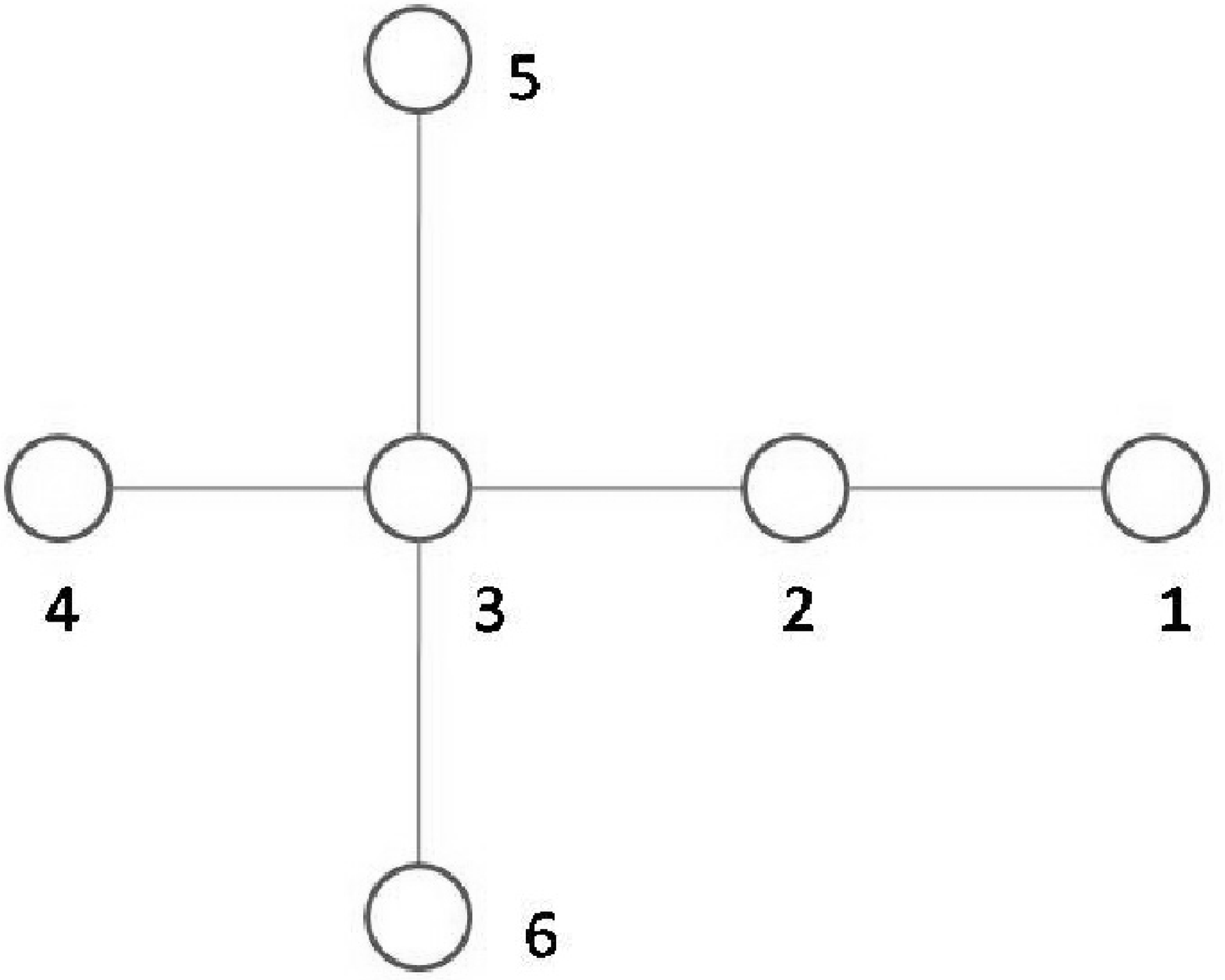}}

\noindent Among several possible choices, our 2 examples lead us
respectively to sub-algebras $ A_4 $ and $ D_5 $ of $H$, in view of
the following choices:

\noindent $ (1) \ \  I_1 = \{\alpha_1, \dots , \alpha_4 \} \subset
S$

\noindent $ (2) \ \  I_2 = \{\alpha_1, \dots , \alpha_5 \} \subset
S$

\noindent We note here that there could be no choice for a subset
which allows us to get a Lie sub-algebra which is not contained
inside the Dynkin diagram of $H$. For the first case,
above-mentioned theorem leads us to a Poincare polynomial
$$ P(H) = P(A_4) \ R_1  \eqno(I.6) $$
where $P(A_4)$ is the Poincare polynomial of $A_4$ Lie algebra with
the Dynkin diagram

\epsfxsize=6cm \centerline{\epsfbox{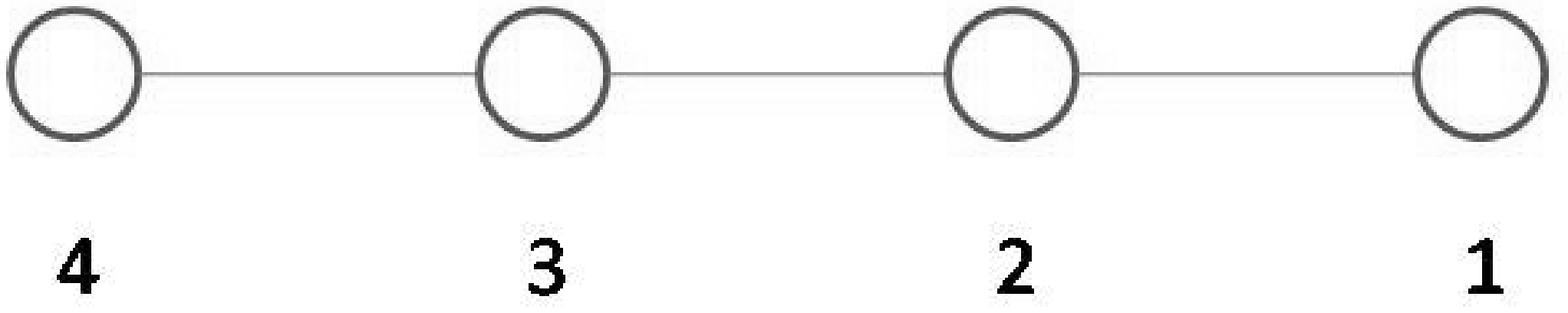}}

\noindent and $R_1$ is a rational function which we calculate here,
explicitly. In favor of (I.1), we know that
$$ P(A_4) =1 + 4 \ t + 9 \ t^2 + 15 \ t^3 + 20 \ t^4 + 22 \ t^5 + 20 \ t^6 + 15 \ t^7 +
 9 \ t^8 + 4 \ t^9 + t^{10}  $$
\noindent and corresponding 120 elements of $ W(H)$ form a subset
$W({A_4}) \subset W(H)$. Among infinite number of elements of $
W(H)$, there are only 120 elements originate only from the set $ I_1
\subset S $ of example (1). In (I.6), $R_1$ originates from a subset
$R_1(H) \subset W(H) $ for which any element $ w \in W(H)$ can be
expressed by the factorized form
$$ w = u \ v \ \ , \ \ u \in W({A_4}) \ \ , \ \ v \in R_1(H)  \eqno(I.7) $$
in such a way that
$$ \it{l}(w)= \it{l}(u) + \it{l}(v)   \eqno(I.8)  $$
where $ \it{l} $ is the length function of $H$. Due to space
limitation, we exemplify our algorithm by giving elements of
$R_1(H)$ up to 6th order:

$$ \eqalign{
 &1 \cr
 &\sigma_5 , \sigma_6 , \cr
 &\Sigma(5,3) ,\Sigma(5,6) , \Sigma(6,3) , \cr
 &\Sigma(5,3,2) , \Sigma(5,3,4) , \Sigma(5,3,6) , \Sigma(5,6,3) , \Sigma(6,3,2) , \Sigma(6,3,4) , \Sigma(6,3,5) \cr
 &\Sigma(5,3,2,1) , \Sigma(5,3,2,4) , \Sigma(5,3,2,6) , \Sigma(5,3,4,6) , \Sigma(5,3,6,3) \Sigma(5,6,3,2) , \cr
 &\Sigma(5,6,3,4) , \Sigma(5,6,3,5) , \Sigma(6,3,2,1) , \Sigma(6,3,2,4) , \Sigma(6,3,2,5) , \Sigma(6,3,4,5) , \cr
 &\Sigma(5,3,2,1,4) , \Sigma(5,3,2,1,6) , \Sigma(5,3,2,4,3) , \Sigma(5,3,2,4,6) , \cr
 &\Sigma(5,3,2,6,3) , \Sigma(5,3,4,6,3) , \Sigma(5,3,6,3,2) , \Sigma(5,3,6,3,4) , \Sigma(5,3,6,3,5) , \cr
 &\Sigma(5,6,3,2,1) , \Sigma(5,6,3,2,4) , \Sigma(5,6,3,2,5) , \Sigma(5,6,3,4,5) , \Sigma(6,3,2,1,4) , \cr
 &\Sigma(6,3,2,1,5) , \Sigma(6,3,2,4,3) , \Sigma(6,3,2,4,5) , \Sigma(6,3,2,5,3) , \Sigma(6,3,4,5,3) , \cr
 &\Sigma(5,3,2,1,4,3) , \Sigma(5,3,2,1,4,6) , \Sigma(5,3,2,1,6,3) , \Sigma(5,3,2,4,3,5) , \cr
 &\Sigma(5,3,2,4,3,6) , \Sigma(5,3,2,4,6,3) , \Sigma(5,3,2,6,3,2) , \Sigma(5,3,2,6,3,4) , \cr
 &\Sigma(5,3,2,6,3,5) , \Sigma(5,3,4,6,3,2) , \Sigma(5,3,4,6,3,4) , \Sigma(5,3,4,6,3,5) , \cr
 &\Sigma(5,3,6,3,2,1) , \Sigma(5,3,6,3,2,4) , \Sigma(5,3,6,3,2,5) , \Sigma(5,3,6,3,4,5) , \cr
 &\Sigma(5,6,3,2,1,4) , \Sigma(5,6,3,2,1,5) , \Sigma(5,6,3,2,4,3) , \Sigma(5,6,3,2,4,5) , \cr
 &\Sigma(5,6,3,2,5,3) , \Sigma(5,6,3,4,5,3) , \Sigma(6,3,2,1,4,3) , \Sigma(6,3,2,1,4,5) , \cr
 &\Sigma(6,3,2,1,5,3) , \Sigma(6,3,2,4,3,5) , \Sigma(6,3,2,4,3,6) , \Sigma(6,3,2,4,5,3) , \cr
 &\Sigma(6,3,2,5,3,4) , \Sigma(6,3,2,5,3,6) , \Sigma(6,3,4,5,3,2) , \Sigma(6,3,4,5,3,6)    } $$
\noindent The reader could verify order by order that the number of
these elements do match with the first 6 terms in the infinite
polynomial expansion of the following rational function:
$$ R_1 ={(1 + t)^3 (1 + t^2) (1 - t + t^2) (1 + t^4) \over
1 - t^2 - 2 t^3 - t^4 + t^6 + t^7 + 3 t^8 + 2 t^9 - t^{13} - 2
t^{14} - 2 t^{15} - t^{16} + t^{19} + t^{20} } $$ \noindent Our
algorithm however allows us to investigate the existence of (I.6) at
any order. To this end, let us consider
$$ P(H) \equiv \sum_{n=0}^\infty w_n \ t^n $$
and
$$ P(A_4) \equiv \sum_{n=0}^{10} u_n \ t^n   . $$
Since $R_1$ is a rational function, it could be represented also by
a polynomial of infinite order:
$$ R_1 \equiv \sum_{n=0}^\infty v_n \ t^n $$
The reader could verify now that at any order $ M=0 , \dots , \infty
$
$$ w_M = \sum_{s=0}^M u_s \ v_{M-s} . \eqno(I.9) $$

In case of example (2), one finds the following Dynkin diagram

\epsfxsize=6cm \centerline{\epsfbox{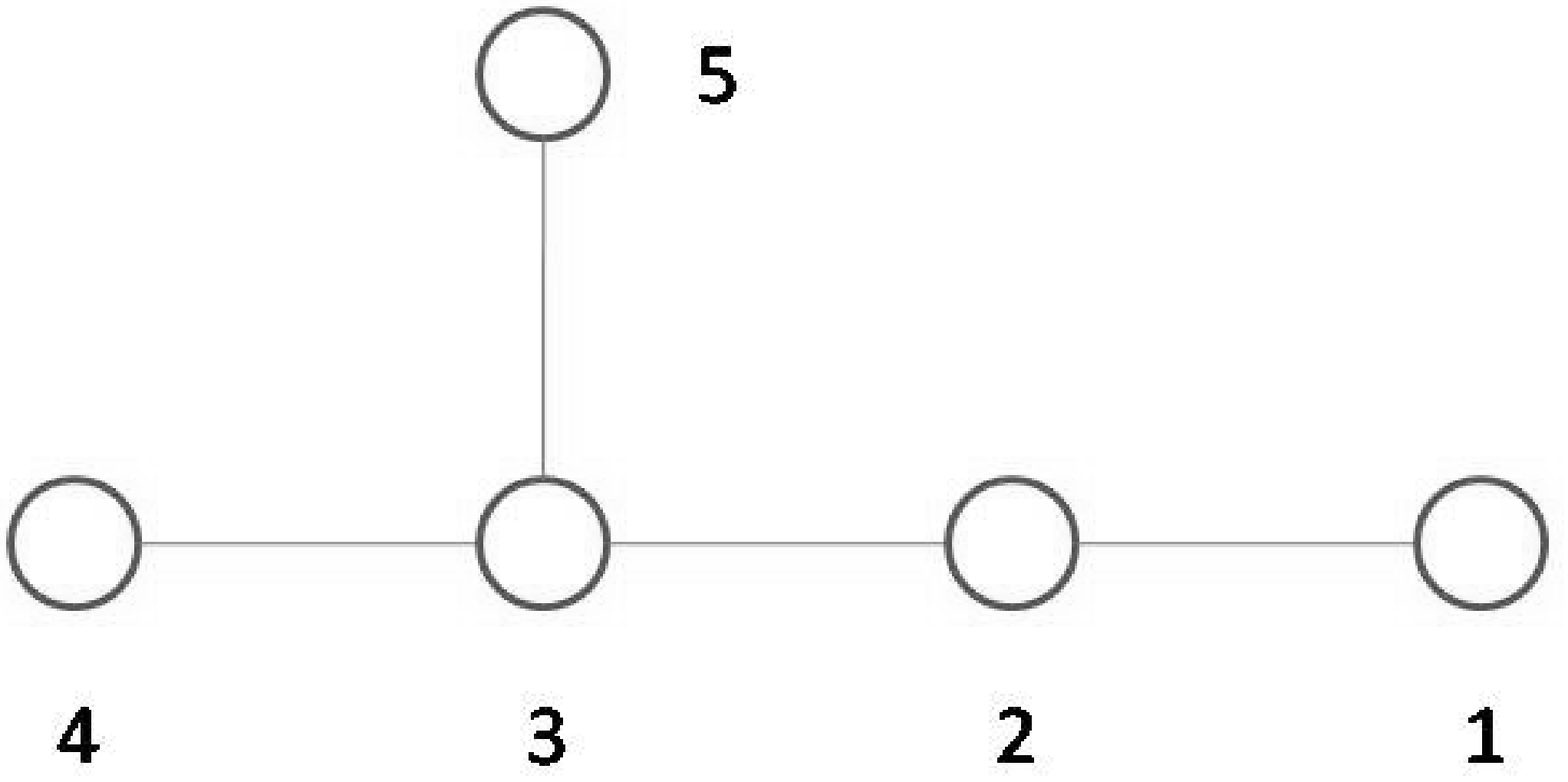}}

\noindent which gives us
$$ P(H) = P(D_5) \ R_2   \eqno(I.10) $$
where $ P(D_5) $ is the Poincare polynomial of $D_5$ Lie algebra. As
in the first example, $R_2$ is to be calculated in the form of the
following rational polynomial :
$$ R_2 ={ (1 + t) \over 1 - t^2 - 2 t^3 - t^4 + t^6 + t^7 + 3 t^8 + 2 t^9 - t^{13} - 2
t^{14} - 2 t^{15} - t^{16} + t^{19} + t^{20}}   $$

\noindent and the similar of (I.9) is seen to be valid.

For our last example, the Dynkin diagram of $H$ should be enumerated
not in the way defined above but as in the following:

\epsfxsize=6cm \centerline{\epsfbox{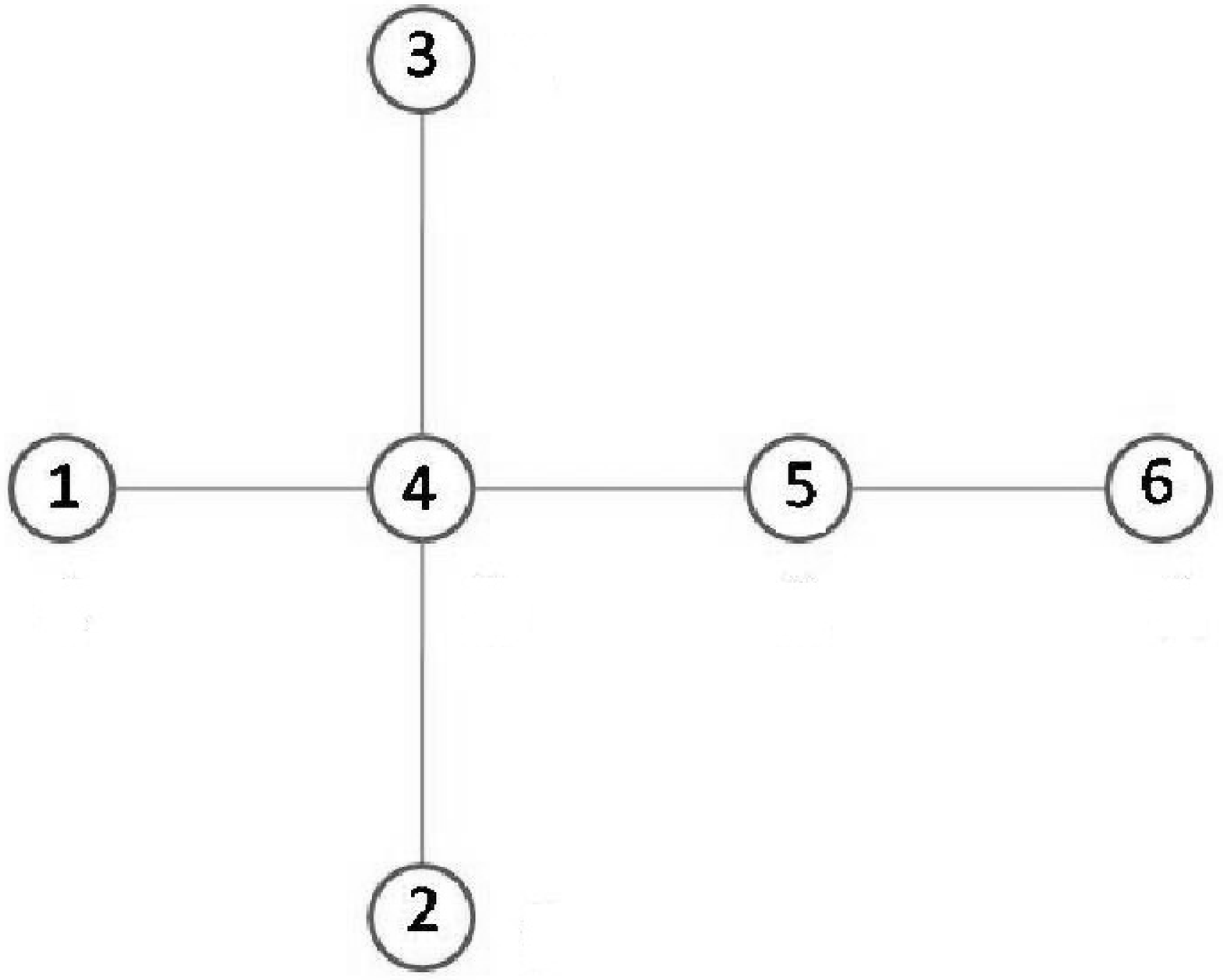}}

\noindent in such a way that the choice \noindent

\noindent $ (3) \ \  I_3 = \{\alpha_1, \dots , \alpha_5 \} \subset
S$

\noindent gives us an infinite sub-algebra which is in fact the
affine Lie algebra $ \widehat{D}_4 $ with the following Dynkin
diagram:

\epsfxsize=5cm \centerline{\epsfbox{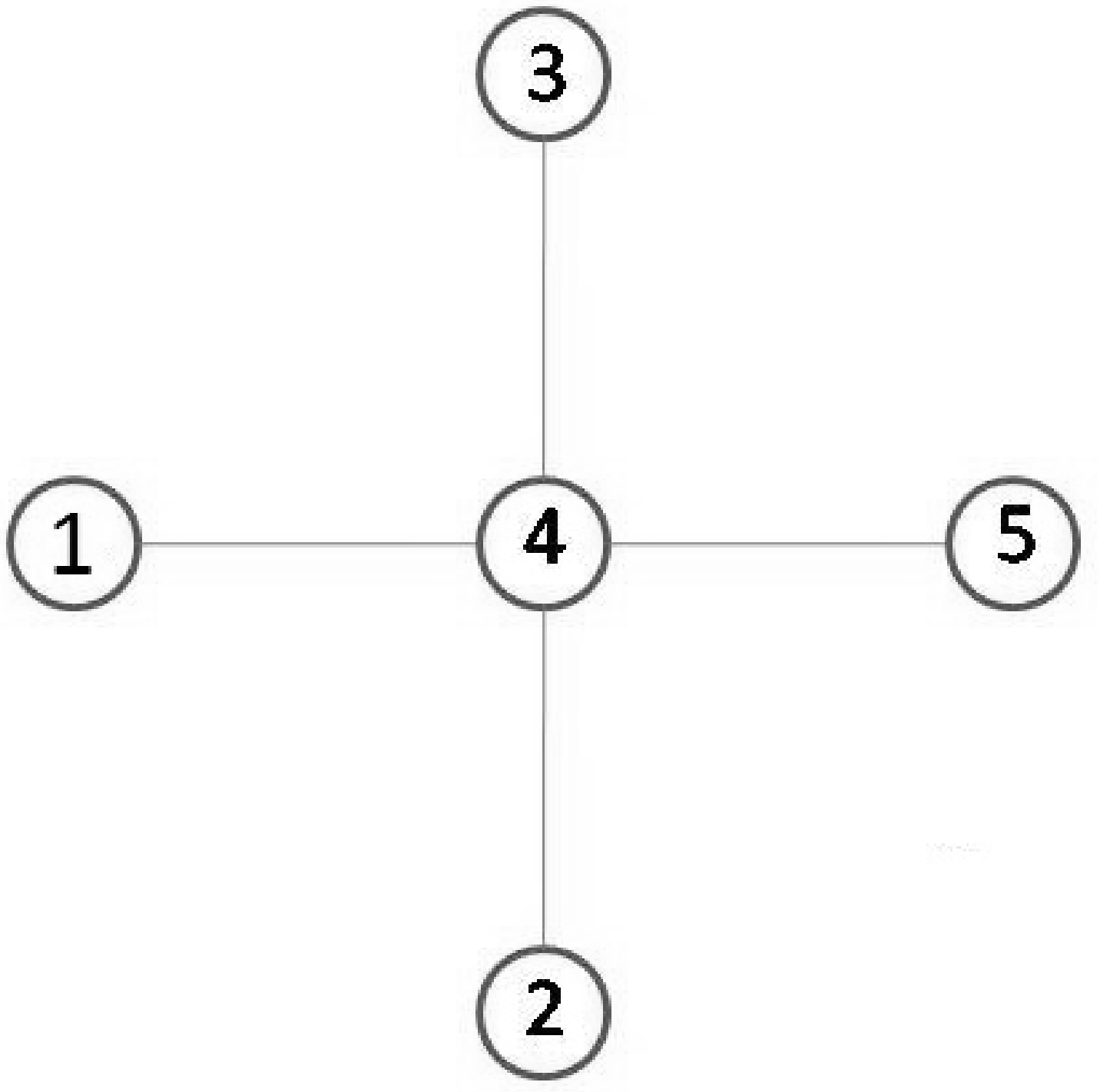}}

\noindent This time, similar to (I.6) and (I.10), one obtains
$$ P(H) = P(\widehat{D}_4) \ R_3   \eqno(I.11) $$
\noindent where
$$ R_3 = {(1 - t)^3 (1 + t) (1 + t + t^2)^2 (1 + t^4) (1 + t + t^2 + t^3 + t^4) \over
 1 - t^2 - 2 t^3 - t^4 + t^5 + t^6 + t^8 + t^9 + t^{10} + t^{11} - t^{14} -
 t^{15} } . $$
\noindent From (I.2), we know that $ P(\widehat{D}_4) \equiv
\sum_{n=0}^\infty u_n \ t^n  $ . In result, we see that (I.9) is
also valid here.

Let us conclude this section by giving the main motivation behind
this work. As the theorem said, in all three of above examples,
corresponding polynomials $R_1, R_2$ and also $R_3$ are rational
functions. In the next section, we show by explicit calculation that
(I.4) is in fact valid for Poincare polynomial of H. From above
discussion, it is seen that theorem can not give a way to obtain
(I.4).


\vskip 3mm \noindent {\bf{II.\ CALCULATION OF HYPERBOLIC POINCARE
POLYNOMIALS }} \vskip 3mm

We follow Humphreys {\bf[5]} for Lie algebraic technology and
Kac-Moody-Wakimoto {\bf [6]} for the description of hyperbolic Lie
algebras. To explain formal structure of our calculations, we follow
the example $ H $ for which simple roots $\alpha_i$ and fundamental
dominant weights $\lambda_i$ are given by
$$ \kappa(\lambda_i,\alpha_j) = \delta_{i,j} \ \ \ \ \ \  i,j = 1, \dots ,6. $$
where $\kappa( ,)$ is the symmetric scalar product which we know to
be exist, on $ H$ weight lattice, by its Cartan matrix A
$$ A = - \pmatrix{
 2&& -1&& 0&& 0&& 0&& 0&& \cr
 -1&& 2&& -1&& 0&& 0&& 0&& \cr
 0&& -1&& 2&& -1&& -1&& -1&& \cr
 0&& 0&& -1&& 2&& 0&& 0&& \cr
 0&& 0&& -1&& 0&& 2&& 0&& \cr
 0&& 0&& -1&& 0&& 0&& 2&& }$$
\noindent so we have
$$ \lambda_{i} = \sum_{j=1}^6 (A^{-1})_{i,j} \ \alpha_j $$

Although following discussions are valid for any $G_N$, we still
proceed in H mentioned above. To this end, let $ W(H)$ be the Weyl
group and $\rho$ be the Weyl vector of H. For any $\Sigma \in W(H)$,
let us now consider
$$ \Gamma \equiv \rho-\Sigma(\rho) \eqno(II.1)$$
which is by definition an element of the positive root lattice of
$H$. We know that $\Gamma$ is unique in the sense that  $\Gamma
\equiv \rho-\Sigma(\rho)$ is different from $\Gamma^\prime \equiv
\rho-\Sigma^\prime(\rho)$ for any two different $\Sigma$,
$\Sigma^\prime \in W(H)$. Note here that the Weyl vector $\rho$ is a
strictly dominant weight. This is sufficient to propose our simple
method to calculate the number of Weyl group elements which are
expressed in terms of the same number of simple Weyl reflections
$\sigma_i$ which are defined by
$$ \sigma_i(\Lambda) \equiv \Lambda - 2 \
{ \kappa(\Lambda,\alpha_i) \over \kappa(\alpha_i,\alpha_i) } \
\alpha_i  \ \ , \ \ i =1,2,3, \dots $$ for any element $\Lambda$ of
weight lattice.

As in above, the k-tuple products $ \Sigma(i_1, \dots ,i_k) $ are
the reduced elements which can not be reduced into products
consisting less than k-number of simple Weyl reflections. Out of all
these reduced elements, we define a class $ W^k \subset W(G_N)$.
Different elements of any class $W^k$ are determined uniquely by
their actions on the Weyl vector. We use a definitive algorithm to
choose the ones among the equivalents. The results of this algorithm
are given in above examples. The aim of this work, however, doesn't
need to show further this algorithm here.

Now we can formally state that a Weyl group is the formal sum of its
classes $ W^k$. One should note that the order $ \mid W^k \mid $ of
a class $ W^k $ is always finite though the number of these classes
is finite for finite and infinite for infinite Kac-Moody Lie
algebras.

Looking back to $H$, we give some of its classes in the following:

$$ \eqalign{
&W^0 = \{ 1 \} \cr &W^1 = \{ \sigma_1, \dots , \sigma_6 \} \cr &W^2
= \{ \Sigma(1,2) , \Sigma(1,3) , \Sigma(1,4) , \Sigma(1,5) ,
\Sigma(1,6) , \cr &\ \ \ \ \ \ \ \ \ \ \Sigma(2,1) ,  \Sigma(2,3) ,
\Sigma(2,4) , \Sigma(2,5) , \Sigma(2,6) , \cr &\ \ \ \ \ \ \ \ \ \
\Sigma(3,2) , \Sigma(3,4) , \Sigma(3,5) , \Sigma(3,6) , \Sigma(4,3)
, \cr &\ \ \ \ \ \ \ \ \ \ \Sigma(4,5) , \Sigma(4,6) , \Sigma(5,3) ,
\Sigma(5,6) , \Sigma(6,3) \} \cr & \dots } $$

\noindent As a result, one has a polynomial $\sum_{k=0}^\infty \
\mid W^k \mid \ t^k $ which is nothing but the Poincare polynomial
of $H$, as is mentioned in sec.I.

By explicit calculation up to 26th order, we obtained the following
result
$$ \eqalign{P(H) &\equiv
1 + 6 \ t^1 + 20 \ t^2 + 52 \ t^3 + 117 \ t^4 + 237 \ t^5 + 445 \
t^6 + 791 \ t^7 + 1347 \ t^8 + 2216 \ t^9 \cr &+ 3550 \ t^{10} +
5568 \ t^{11} + 8582 \ t^{12} + 13044 \ t^{13} + 19604 \ t^{14} +
29189 \ t^{15} \cr &+  43129 \ t^{16} + 63332 \ t^{17} + 92518 \
t^{18} + 134572 \ t^{19} + 195052 \ t^{20} + 281882 \ t^{21} \cr &+
406361 \ t^{22} + 584620 \ t^{23} + 839655 \ t^{24} + 1204232 \
t^{25} + \dots } \eqno(II.2)
$$ One sees that (II.2) is sufficient to conclude that
$$ P(H) \equiv { P(B_5) \over Q(B_5) } \eqno(II.3) $$
where
$$ Q(B_5) \equiv (1 - t - 2 \ t^3 + t^4 + t^6 - t^7 + 2 \
t^8 - t^9 + t^{10} + t^{12} + t^{13} - t^{14} - t^{15} -t^{18} -
t^{20} + t^{24}) \eqno(II.4) $$ and $P(B_5)$ comes from (I.1) for
$B_5$ Lie algebra. Note here that the number of positive roots of
$B_5$ is equal to D=25 and hence $Q(B_5)$ is a polynomial of order
D-1=24.

\vskip 3mm \noindent {\bf{III.\ CONCLUSION }} \vskip 3mm

Let us conclude with the main idea of this work by the aid of a
beautiful example. One could say, for instance, that Bott theorem
and also the factorization theorem given above say the same thing.
Although this theorem proves useful as a calculational tool, Bott
theorem gives us a general framework to apply for affine Lie
algebras due to the fact that explicit calculation of a rational
function is in fact quite hard if it is not impossible. We note
again that a rational function can be expressed only in the form of
a polynomial of infinite order.

In the lack of such a general framework for Lie algebras beyond
affine ones, we also use an algorithm for explicit calculations.
Against the above-mentioned theorem, explicit calculations are
possible here due to the fact that in our formalism we only deal
with polynomials of some finite degree.

\vskip3mm \noindent{\bf {REFERENCES}} \vskip3mm

\item [1] J. E. Humphreys, Reflection Groups and Coxeter Groups,
\item \ \ \ \ \ Cambridge University Press, 1990
\item [2] R. Bott, An Application of the Morse Theory to the topology of Lie-groups,
\item \ \ \ \ \ Bull. Soc. Math. France 84 (1956) 251-281
\item [3] N. Bourbaki, Groupes et Algebres de Lie, Chapter 4-6,
Masson, Paris, 1981
\item \ \ \ \ \ and also p.123 of [1].
\item [4] V. Kac, Infinite Dimensional Lie Algebras, Cambridge University Press, 1982
\item [5] J. E. Humphreys, Introduction to Lie Algebras and Representation Theory,
\item \ \ \ \ Springer-Verlag, 1972
\item [6] V. G. Kac, R. V. Moody and M. Wakimoto, On $E_{10}$, Differential Geometrical
\item \ \ \ \ Methods in Theoretical Physics, Kluwer Acad. Pub., 1988

\end